\documentstyle[psfig,twocolumn,prl,aps,amssymb]{revtex}
\begin{document}
\wideabs{
\draft
\title{Even-Odd Effect in Spontaneously Coherent Bilayer Quantum Hall Droplets}
\author{K. Park, V. W. Scarola, and S. Das Sarma}
\address{Condensed Matter Theory Center and 
Department of Physics, University of Maryland,
College Park, MD 20742-4111}
\date{\today}

\maketitle

\begin{abstract}
Using exact diagonalization in the disc geometry
we predict a novel even-odd effect 
in the Coulomb blockade spectra of
vertically coupled double quantum dots
under an external magnetic field.
The even-odd effect in the tunneling conductance
is a direct manifestation of spontaneous interlayer phase coherence,
and is similar to the even-odd resonance in the Cooper pair box problem
in mesoscopic superconducting grains.
Coherent fluctuations in the number of Cooper pairs 
in superconductors is analogous to
the fluctuations in the relative number difference 
between the two layers in quantum Hall droplets.
\end{abstract}

\pacs{PACS numbers:}}

It is now well accepted \cite{Review} that a bilayer quantum Hall
system may spontaneously develop interaction-induced
interlayer phase coherence with an associated Goldstone mode.
Such an interlayer coherent state is akin to an excitonic condensate,
being qualitatively similar to a neutral superfluid ground state.
Pioneering experiments \cite{Eisenstein} by Eisenstein and collaborators
have firmly established the physical reality of such an interlayer
coherent phase in high-mobility bilayer GaAs heterostructures
around a total Landau level filling factor of unity ($\nu_T =1$).
One of the most spectacular experimental demonstrations of 
the superfluidity of this spontaneous coherent phase has been
the observation \cite{Eisenstein} of a very sharp interlayer tunneling peak
which has been interpreted by some 
(but not all) 
as the direct analog of the Josephson effect \cite{Review}.

Given the considerable significance of the spontaneous quantum Hall
interlayer phase coherence as a novel correlation-induced collective
phenomenon,
it is important to envisage alternative non-trivial
properties of the coherent state which have direct analogies to
superconducting systems.
In this Letter, we theoretically study one such property,
namely the precise analogy between bilayer coherent quantum Hall
droplets and the ``Cooper pair box'' problem in small superconducting 
grains studied in a series of seminal experiments \cite{Nakamura}
by Nakamura and collaborators. 
Our theoretical results presented in this paper shows convincingly
that bilayer quantum Hall quantum dot systems could
spontaneously (i.e. in the absence of any interlayer tunneling)
develop interlayer coherence leading to coherent fluctuations
in the number of electrons in each dot (with the total number of
electrons in the double dot system being fixed),
which in turn will give rise to an experimentally observable even-odd 
effect analogous to the even-odd resonance
(and the corresponding Rabi oscillations) reported
in the Cooper pair box experiment \cite{Nakamura}.
Our predicted even-odd effect in the bilayer quantum Hall quantum dot
system has the obvious additional exciting prospect of providing a
robust quantum two-level system
with the interesting potential of serving as a qubit in a novel
quantum-Hall-quantum-dot quantum computer architecture \cite{Yang},
which is fundamentally different from electron-spin-based
quantum dot qubits currently being studied in the literature
\cite{QC}.

The system of interest is the so-called
vertically-coupled double dot system \cite{Tarucha}
in a strong, external magnetic field so that
the double-dot system is effectively equivalent to a finite
bilayer quantum Hall droplet. We also assume that the
electron number in each dot can be precisely controlled
as has been demonstrated.
In particular, we consider the system to have an odd number of
electrons,
which, to be specific, we take to be 7 without any loss of generality
(any other small odd number of electrons such as 5, 9, 11, 13, etc.
does not make any difference in our analysis).
Due to the large capacitive energy associated
with interlayer charge imbalance,
the ground state of the system
(without any external bias voltages and interlayer tunneling)
has double degeneracy: the state with 4 (3) electrons
in the top (bottom) layer has exactly the same energy as
the state with 3 (4) electrons in the top (bottom) layer.
We denote these two degenerate ground states as 
$|4,3\rangle$ and $|3,4\rangle$, respectively.

For a range of magnetic fields
the two degenerate ground states, 
$|4,3\rangle$ and $|3,4\rangle$, 
are separated from 
the excitation spectrum 
by an energy gap.
This is so because 
the Coulomb interaction energy 
has a {\it cusp} 
at a particular configuration of states
which is usually known as 
the maximum density droplet (MDD) state \cite{MDD}.
It is important to note that
the MDD state in our system
is the mesoscopic droplet realization of 
the bulk bilayer coherent state at $\nu_T=1$.
The above two MDD states,
$|4,3\rangle$ and $|3,4\rangle$, 
compose our two level system for the finite droplet
similar to the corresponding Cooper pair box situation.

We begin our quantitative analysis
by considering the Hamiltonian for electrons 
subject to a uniform magnetic field and 
a parabolic confining potential.
In the Fock-Darwin basis,
the many-body Hamiltonian for the bilayer quantum dot system
can be written as:
\begin{eqnarray}
H &=& H_0 + \hat{P} V_{Coul} \hat{P} + H_{t}
\label{Hamiltonian}
\end{eqnarray}
where 
$H_0 = \frac{1}{2} 
\left( \sqrt{\omega^2_c +4\omega^2_0}
-\omega_c \right) \hat{L}_z $,
with $\hat{L}_z$ being the total angular momentum in $z$-direction. 
Also, $\omega_c$ is the cyclotron frequency and 
$\omega_0$ parameterizes the parabolic confining potential.  
$\hat{P}$ is the lowest Landau level (LLL) projection operator and 
$V_{Coul}$ represents the usual Coulomb interaction
between electrons:
\begin{eqnarray}
\frac{V_{Coul}}{e^2/\epsilon a} = \sum_{i<j \in \uparrow} \frac{1}{r_{ij}} 
+\sum_{k<l \in \downarrow} \frac{1}{r_{kl}}
+ \sum_{i\in \uparrow,k\in\downarrow} 
\frac{1}{\sqrt{r^2_{ik}+(\frac{d}{a})^2}},
\end{eqnarray}
where  $d$ is the interlayer spacing, $\epsilon$ is the 
GaAs dielectric constant,
and $r_{ij}$ is the lateral separation between
the $i$-th and $j$-th electron.  
The natural length unit
is the modified magnetic length 
$a=l_B (1+4 \omega_0^2/\omega_c^2)^{-1/4}$ which 
reduces to the planar 
magnetic length, $l_B =\sqrt{\hbar c/e B}$, when the 
cyclotron energy is much larger than the confining potential energy.
In the above we have used a 
pseudo-spin representation to describe
the double layer system: 
$\uparrow$ and $\downarrow$ distinguish different layers. 
In general we define the pseudo-spin operator:
\begin{eqnarray}
{\bf S} \equiv 
\frac{1}{2} \sum_{m} c^{\dagger}_{a}(m) \vec{\sigma}_{ab} c_{b}(m),
\end{eqnarray}
where $\hat{S}_z$ measures 
the electron number difference between layers,
and $\hat{S}_x$ is associated with interlayer tunneling.
We take the real spin to be
fully polarized 
either because of the large Zeeman coupling
or because of electron-electron repulsion
, i.e. Hund's rule.

The tunneling Hamiltonian $H_{t}$ in the Eq.(\ref{Hamiltonian})
can be written as:
\begin{eqnarray}
H_{t} = -\frac{t}{2} \sum_{m} c^{\dagger}_{a}(m) \sigma_{ab}^x c_{b}(m)
\equiv -t \hat{S}_x  
\label{H_t}
\end{eqnarray}
where $\sigma$ is the usual Pauli matrix,  
$t$ is the single particle interlayer tunneling gap,
and $m$ denotes the LLL angular momentum quantum number.
Eq.(\ref{H_t}) is valid for general $t$.  However,
we are interested in the limit 
of zero interlayer tunneling, 
i.e. $t/(e^2/\epsilon a) \rightarrow 0$,
which is appropriate when 
considering \emph{spontaneous} interlayer coherence
(note that the $t \rightarrow 0$ limit is not 
the same as the $t=0$ situation).

We now analyze the Hamiltonian in Eq.(\ref{Hamiltonian})
via exact diagonalization of finite size systems 
in the limit of zero tunneling. 
In this limit, the Hamiltonian 
is invariant with respect to layer 
exchange and spatial rotations,
i.e. $[V_{Coul},\hat{S}_z]=0$ and $[V_{Coul},\hat{L}_z]=0$.
We may therefore restrict the Hilbert space to 
specific $S_z$ and $M_z$ sectors, where $S_z$ and $M_z$ are the 
eigenvalues of $\hat{S}_z$ and $\hat{L}_z$, respectively.
We diagonalize the Coulomb interaction in the 
basis of LLL single-particle eigenstates. \cite{Basis}. 
In particular,
we focus our attention 
on the part of the Hilbert space containing the MDD state
which occurs at $M_z = N(N-1)/2$, where $N$ is the total 
number of particles.  

\begin{figure}
\centerline{\psfig{figure=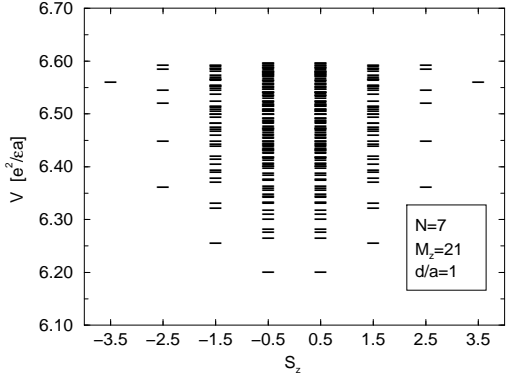,width=3.5in,angle=-90}}
\caption{Coulomb interaction energy as a function of 
$S_z$ which is half the
relative electron number difference between different layers.
The layer separation is chosen to be $d/a=1$ where 
we define 
$a=l_B (1+4 \omega_0^2/\omega_c^2)^{-1/4}$,
the magnetic length $l_B=\sqrt{\hbar c/eB}$,
and the cyclotron frequency $\omega_c = eB/m^* c$.  
$\omega_0$ is the frequency of the confining potential.
The state with angular momentum $M_z=21$,
is the maximum density droplet state for $N=7$.
\label{fig1}}
\end{figure}

Fig.\ref{fig1} shows the eigenenergy spectrum 
of the Coulomb interaction as a function of $S_z$ 
for a 7 electron system with $d/a=1$ and $M_z = 21$.   
Due to the direct electrostatic 
contribution of the Coulomb interaction,
the lowest energy state is obtained for states 
with the smallest charge imbalance 
between layers,
i.e. $S_z=+1/2$ ($|4,3\rangle$) and $-1/2$ ($|3,4\rangle$).
In fact, this electrostatic contribution 
may be viewed as the relative charging energy
between layers.
As expected, the ground states 
located at $S_z = 1/2$ and $-1/2$ 
are separated from the lowest energy 
states of higher $|S_z|$ 
by the relative charging energy cost:
\begin{eqnarray}
V^{direct}_{Coul} = \frac{\alpha}{N} \hat{S}^2_z,
\end{eqnarray}
where we find 
$\alpha/(e^2/\epsilon a) \simeq -0.18 + 0.35 d/a$ 
for $d/a \gtrsim 0.5$.
In particular, the lowest energy state of $S_z=\pm1/2$
is separated from that of $S_z=\pm3/2$ by
an energy gap of roughly 0.05 $e^2/\epsilon a$.
Therefore,
as far as excitations lower than 
this charging energy cost are concerned,
we can restrict our attention to the Hilbert space
of $S_z = \pm 1/2$. 
Note that the energy spectra of
the $S_z = 1/2$ and $-1/2$ states
are identical because of reflection symmetry 
between layers.
Also, it is important to remember that 
the relative charging energy,
$V^{direct}_{Coul}$,
is inversely proportional to the number of electrons.
Therefore, for large $N$, mixing between states with different 
$S_z$ becomes appeciable, in which case our 
two-level system is ill-defined.

\begin{figure}
\centerline{\psfig{figure=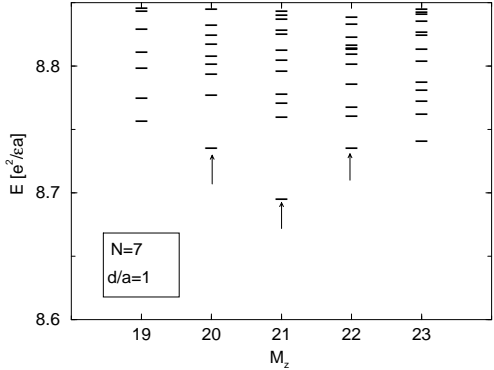,width=3.5in,angle=-90}}
\caption{Energy spectrum as a function of 
total angular momentum, $M_z$,
in the Hilbert space of $S_z = 1/2$.
Note that the energy in the graph
is a sum of the Coulomb interaction energy and
the confining potential energy:
$E = V_{Coul} + \gamma M_z$ where
$\gamma = \frac{\hbar}{2} \left( \sqrt{\omega^2_c +4\omega^2_0}
-\omega_c \right)$.
In the graph, we have chosen $\gamma = 0.1187 e^2/\epsilon a$
which gives us the largest possible gap.  
The arrows indicate the three lowest energies.
\label{fig2}}
\end{figure}

\begin{figure}
\centerline{\psfig{figure=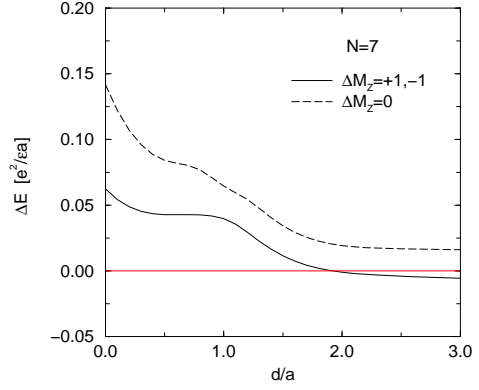,width=3.5in,angle=-90}}
\caption{The lowest three energy gaps as a function of
interlayer separation $d/a$. 
The lowest three excitations are categorized as follows:
(i) $\Delta M_z = +1$ (the edge excitation),
(ii) $\Delta M_z = 0$ and
(iii) $\Delta M_z = -1$ (the internal excitation).
The energy gap is given by:
$\Delta E= \Delta V_{Coul} +\gamma \Delta M_Z$
where $\gamma$ is defined as in Fig. \ref{fig2}.
\label{fig3}}
\end{figure}

Fig.\ref{fig2} shows the energy spectrum in the $S_z=1/2$
Hilbert space as a function of $M_z$ for 7
electrons at $d/a=1$.
The energy in the graph
is the sum of the
Coulomb interaction energy and 
the confining potential energy: 
$E = V_{Coul} + \gamma M_z$ where
$\gamma = \frac{\hbar}{2} \left( \sqrt{\omega^2_c +4\omega^2_0}
-\omega_c \right)$.
By choosing $\gamma = 0.1187 e^2/\epsilon a$
we obtain the maximum gap.
The MDD state is separated 
from the edge excitation ($\Delta M_z = +1$)
and the internal excitation ($\Delta M_z = -1$)
by roughly $0.05 e^2/\epsilon a$
at the interlayer separation $d/a = 1$.
Also, the $\Delta M_z=0$ excitation is shown to have
an energy gap roughly equal to $0.07 e^2/\epsilon a$
at $d/a = 1$.
Fig.\ref{fig3} plots the lowest energy gaps 
as a function of $d/a$.
As seen from the graph, 
the energy gap is well developed for $d/a \lesssim 1$.
We conclude that
the MDD state is stabilized 
in a suitable range of magnetic fields and interlayer distances 
for small system sizes.

\begin{figure}
\centerline{\psfig{figure=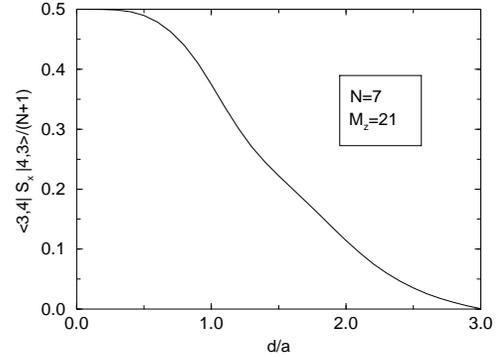,width=3.5in,angle=-90}}
\caption{Interlayer coherence in the limit of zero tunneling
as a function of interlayer separation $d/a$. 
$| 4,3 \rangle$ 
represents the lowest energy state with 
4 (3) electrons in the top (bottom) layer
for a seven electron system, which can be alternatively
denoted by $|S_z=+1/2\rangle$.
$|3,4\rangle$ ($|S_z= -1/2 \rangle$) is similarly defined.
\label{fig4}}
\end{figure}

Now that the two degenerate ground states,
$|S_z=+1/2\rangle$ and $|S_z=-1/2\rangle$,
are shown to be well separated from other excitations
in the limit of zero tunneling,
we can reduce the whole Hilbert space 
into the Hilbert space composed of only these two states. 
In this limit,
the {\it reduced} Hamiltonian is written as:
\begin{eqnarray}
H_{red} = - \Delta_x \sigma_x +\Delta_z \sigma_z,
\end{eqnarray}
where $\sigma_x$ and $\sigma_z$ are the usual Pauli matrices.
In the limit of a small single-particle tunneling gap $t$,
$\Delta_x = t \langle +1/2 | \hat{S}_x | -1/2 \rangle$. 
Also,
$\Delta_z$ is the relative bias voltage between layers.
$\Delta_x$ is the {\it renormalized tunneling gap}
which is greatly enhanced
from the single-particle tunneling gap, $t$, 
by the Coulomb interaction.
In other words, $\Delta_x / t$ is the natural 
order parameter quantifying 
spontaneous coherence in bilayer quantum Hall systems.
The precise definition of spontaneous phase coherence in
our quantum dot system is given by:
\begin{eqnarray}
\lim_{t\rightarrow 0} \frac{\Delta_x}{t} =
\lim_{t\rightarrow 0} \langle +1/2 | \hat{S}_x | -1/2 \rangle
\neq 0.
\end{eqnarray}
Fig.\ref{fig4} shows $\Delta_x/t$ for a 7 particle 
system as a function of $d/a$.  
We see that the interaction-induced coherence effect is 
sizable for $d/a \lesssim 1$. 
It is important to note that $\Delta_x$ 
increases with system size, more precisely: 
$\Delta_x \simeq  \frac{t}{2} (N+1)$ for small $d/a$.

We have established that
the bilayer quantum Hall droplet
is a natural two level system with intrinsic coherence.
Now we predict 
the {\it even-odd} effect in tunneling conductance,
which can be used for experimental confirmation of coherence.  
Tunneling conductance measurements 
in quantum dot systems in the Coulomb-blockade regime find 
conductance peaks 
when the gate voltage $V_g$
is tuned so that
the total energy of the $N$ electron system
becomes identical to that of the $N+1$ electron system.
The total energy of the bilayer quantum Hall quantum dot system
includes the total charging energy cost, 
which is given by:
\begin{eqnarray}
H_{charging} = \frac{e^2}{2C} \left( N - \frac{C V_g}{e} \right)^2
\end{eqnarray}
where $C$ is the total capacitance of the double dot system.

\begin{figure}
\centerline{\psfig{figure=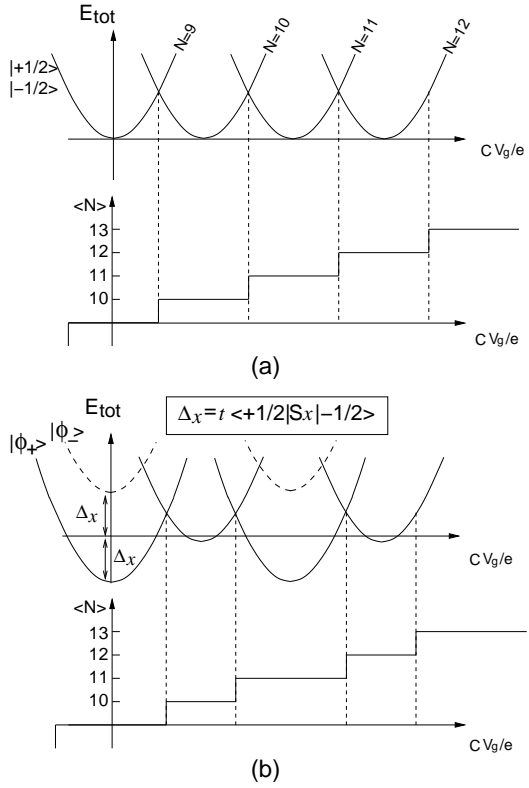,width=3.2in,angle=0}}
\caption{Schematic diagram 
illustrating the even-odd effect
in bilayer quantum Hall dots.
The total energy including the charging energy
is plotted as a function
of $C V_g /e$ where $V_g$ is the gate voltage
and $C$ is the capacitance of the whole bilayer system.
$\langle N \rangle$ is the average number of electrons 
inside double quantum dots.
Fig.\ref{fig5}(a) depicts 
the situation where there is no interlayer coherence. 
$|+1/2\rangle$  ( $|-1/2\rangle$ )
represents the degenerate set of low energy states with 
$S_z =1/2$ ($-1/2$).
It is shown in Fig.\ref{fig5}(b) that, due to the interlayer coherence,
the odd $N$ system acquires an energy splitting
between the symmetric ($|\phi_+\rangle$)
and antisymmetric ($|\phi_-\rangle$)
superposition of
$|+1/2\rangle$ and $|-1/2\rangle$ ,
which is $\Delta_x$.
\label{fig5}}
\end{figure}

Fig.\ref{fig5} shows
a schematic diagram illustrating the even-odd effect.
Fig.\ref{fig5}(a) depicts the case
with no interlayer coherence due to 
either a large interlayer separation or
an extremely small single-particle tunneling gap.
Fig.\ref{fig5}(b) shows that, in the low tunneling limit, 
due to interlayer coherence,
the odd $N$ system acquires an energy splitting
between the symmetric ($|\phi_+\rangle$)
and antisymmetric ($|\phi_-\rangle$)
superposition of
$|+1/2\rangle$ and $|-1/2\rangle$,
which is $\Delta_x$.
On the other hand, the ground state energy of the even $N$ system
decreases by only $2 \tilde{\Delta}^2_x / E_c$ for small $t$,
where $\tilde{\Delta}_x = t \langle 0 |\hat{S}_x| 1 \rangle$,
and $E_c$ is the energy difference 
between $|S_z =0 \rangle$ and $|S_z = 1 \rangle$.
As a result, the distance between conductance peaks 
will oscillate 
between $e^2/C +(\Delta_x -2 \tilde{\Delta}^2_x / E_c)$
and $e^2/C -(\Delta_x -2 \tilde{\Delta}^2_x / E_c)$
as a function of $e V_g$.  
A large tunneling gap will eventually destroy 
the even-odd effect
because all electrons will occupy 
the interlayer-symmetric state.
This will mix states with different $S_z$, thereby
destroying our two-level system.
Both large and zero tunneling will therefore lead  
to evenly spaced Coulomb blockade peaks
whereas the even-odd effect will show up for weak (but finite)
tunneling.

We emphasize that our predicted even-odd effect is 
the precise quantum Hall analog
of the Josephson effect in the Cooper pair box problem.
It is important to note that 
the number of Cooper pairs in superconductors 
can be formally mapped to $S_z$
in our coherent bilayer quantum Hall system.
Therefore, the coherent linear combination of 
superconducting states with different numbers of Cooper pairs 
,which leads to the Josephson effect,
is precisely analogous to 
the linear combination of bilayer quantum Hall states 
with different $S_z$,
which is the origin of our even-odd effect.

In conclusion,
we have proposed an even-odd effect in tunneling conductance
through veritcally coupled double quantum dots
in the quantum Hall regime,
which is the direct analog of the Josephson effect in
mesoscopic superconducting grains.

This work is supported by ARDA and LPS.



\end{document}